\documentclass{article}
\usepackage{latexsym,amsmath, amssymb}
\newcounter{draft} 
\newenvironment{proof}[1]{{\em Proof#1:}}{$\Box$}
\newenvironment{acknowledgement}{{\bf Acknowledgements:}}{}
\newcommand{\new}{\newcommand}
\newcounter{letter}

\new{\Nabla}{\bigtriangledown}
\new{\Ricci}{\mathrm{Ricci}}
\new{\Poly}{\mathrm{Poly}}
\new{\Path}{\mathrm{Path}}
\new{\boxtensor}{{\odot}}
\new{\iso}{\cong}
\new{\vol}{\operatorname{Vol}}
\new{\End}{\mathrm{End}}
\new{\id}{\mathbf{1}}
\new{\tr}{\mathrm{tr}}
\new{\op}{\mathrm{op}}
\new{\scalar}{{\mathfrak{r}}}
\new{\NN}{\mathbb{N}}
\new{\RR}{\mathbb{R}}
\new{\LC}{\mathrm{LC}}
\new{\GK}{\mathrm{GK}}
\new{\HK}{\mathrm{HK}}
\new{\MQ}{\mathrm{K}^{\mathrm{qm}}}
\new{\MQO}{{\mathfrak{K}^{\mathrm{qm}}}}
\new{\s}{{\mathfrak{s}}}
\new{\gap}{{\;}}
\new{\triplegap}{{\;\;\;\;\;}}
\new{\doublegap}{{\;\;\;}}
\new{\Opoly}{{\mathcal{O}}}
\new{\dd}{{\mathfrak{d}}}
\new{\tensor}{\otimes}
\new{\bracket}[1]{\left\langle #1 \right \rangle}
\new{\dx}{\psi}
\new{\dy}{\chi}
\new{\dz}{\xi}
\new{\pt}[1]{#1^{||}}
\new{\opdx}[1]{\dx^{#1}}
\new{\oppartialx}[1]{\iota_{#1}}
\new{\bint}{\oint}
\new{\spin}{\mbox{Spin}}
\new{\pfend}{\noindent \hfill $\Box$}
\new{\xsi}{i}
\new{\para}[2]{\subsection{#1} 
\label{par:#2} \ifnum \thedraft=1 \marginpar{\scriptsize{par:#2}} \fi}
\new{\be}{\begin{equation}  
}
\new{\ee}[1]{%
\label{eq:#1}
 \end{equation} \ifnum \thedraft=1 \marginpar{\scriptsize{\em{eq:#1}}}  
\fi \noindent%
}

\new{\eqa}[2]{\begin{align}
#2 \label{eq:#1}
\end{align}
\ifnum \thedraft=1 \marginpar{\scriptsize{\em{eq:#1}}}  \fi
}
\new{\dlabel}[1]{\ifnum \thedraft=1 \marginpar{\scriptsize{\em{#1}}}  \fi}
 \numberwithin{equation}{section}
\newtheorem{theorem}{Theorem}[subsection]

\newtheorem{corollary}{Corollary}[subsection]
\newenvironment{remark}{{\em Remark }}{}
\newtheorem{lemma}{Lemma}[subsection]

\begin{document}

\section*{A Rigorous Path Integral for Supersymmetic Quantum
Mechanics and the Heat Kernel}

 \vfill

\ifnum \thedraft=1 \today  \fi
\begin{centering}\hfill Dana S. Fine\footnote{University of Massachusetts Dartmouth,
  {\em dfine@umassd.edu}}  and Stephen F. Sawin\footnote{Fairfield University,
  {\em sawin@cs.fairfield.edu}} \hfill \end{centering}

\vfill

\begin{abstract}
  In a rigorous construction of the path integral for supersymmetric
  quantum mechanics on a Riemann manifold, based on B\"ar and
  Pf\"affle's use of piecewise geodesic paths, the kernel of the time
  evolution operator is the heat kernel for the Laplacian on
  forms. The path integral is approximated by the integral of a form
  on the space of piecewise geodesic paths which is the pullback by a
  natural section of Mathai and Quillen's Thom form of a bundle over
  this space.  In the case of closed paths, the bundle is the tangent
  space to the space of geodesic paths, and the integral of this form
  passes in the limit to the supertrace of the heat kernel.

\end{abstract}
\vfill
\newpage
\section*{Introduction}

In \cite{BP07} B\"ar and Pf\"affle construct a path integral
representation of the heat kernel for a general Laplacian on a Riemann
manifold. They express the path integral as an integral over piecewise
geodesic paths in the limit as $n$, the number of pieces, approaches
infinity.  In this note, we begin with the Lagrangian for $N=1$
supersymmetric quantum mechanics (SUSYQM), restrict the action to
piecewise geodesic paths, and identify the resulting expression as a
form on a finite-dimensional manifold. 
This form derives directly from
Mathai and Quillen's universal Thom form.  We interpret the integral
of the top part of this form over the finite-dimensional space as
defining an approximation to the path integral representing the kernel
of the SUSYQM time evolution operator.  
Applying B\"ar and Pf\"affle's
arguments to evaluate the appropriate large-$n$ limit shows the
partition functions for piecewise geodesic paths with fixed endpoints
converge to the heat kernel for the Laplacian on forms. Precisely, we
prove a corollary to B\"ar and Pf\"affle's Theorems~2.8 and~6.1:

\setcounter{section}{3}
\setcounter{subsection}{5}
\begin{corollary}\label{par:introcr}
For any sequence of partitions  $t_1, t_2, \ldots,
t_n$ such that  $\max_i (t_i) \to 0$ and $\sum_i t_i \to t$ and for any form
$\alpha$ on $M$
\[\lim {\mathfrak{K}}(t_1) {\mathfrak{K}}(t_2) \cdots {\mathfrak{K}}(t_n) \alpha = e^{-t\Delta/2}
\alpha\]
where $\Delta$ is the Laplace-Beltrami operator on forms.  
Moreover, for some such sequence of partitions 
\[\lim K(t_1) * K(t_2) * \cdots * K(t_n) \to K_\Delta(x,y;t)\]
uniformly, where $K_\Delta$ is the heat kernel of $\Delta $ (the
kernel of $e^{-t\Delta/2}$). 
\end{corollary}
Here the kernel $K(t)$ of the operator ${\mathfrak{K}}(t)$  is the
pullback (by a certain natural section) of Mathai and Quillen's Thom form on the bundle $TM \times M
\to M \times M$ restricted to an open subset. 
In fact, the indicated
$n$-fold $*$-product expresses an integration of the analogous
Mathai-Quillen Thom form on a bundle over
$M^{n+1}$ restricted to an open subset and pulled back by a
section. The base space of this bundle fibers further to become a
bundle over $M \times M$, on which the $n$-fold $*$-product becomes an
integration over the fibers.

\setcounter{section}{0}
\setcounter{subsection}{0}

The import of this corollary is that the  finite-dimensional partition
functions  which  directly
approximate the kernel of 
the time evolution operator $e^{-t \Delta /2}$ converge to the heat kernel.  Further, for
closed paths based at a given point, this yields a rigorous
path integral expression for the supertrace of the heat
kernel. This path integral is the 
large-$n$ limit of the Mathai-Quillen Euler form integrated over the
finite-dimensional manifold.

Getzler~\cite{G86} uses stochastic integrals due to
Stroock~\cite{S70}, and asymptotics of the heat operator for the
Laplacian on spinors due to Patodi~\cite{P71}, to calculate the
supertrace of this heat operator as a rigorous path integral. 
Rogers~\cite{Rogers92} uses
stochastic analysis techniques to express the heat operator on forms in terms of
a supersymmetric generalization of Wiener integrals and thereby
obtains a path integral 
expression for the supertrace of the heat operator. The novelty of our
approach is in constructing a rigorous path integral that directly
links the heat operator to the SUSYQM time
evolution operator and the Mathai-Quillen construction.

These result confirm Alvarez-Gaum\'e's~\cite{Alvarez83} and Witten's~\cite{Witten82} now-standard
arguments, which express the
supertrace of the heat operator heuristically as a path integral. Our approach
to rigorizing these arguments is sufficiently direct to see the
relation, as derived formally by Blau~\cite{Blau93},
between SUSYQM and  Mathai \& Quillen's universal Thom form~\cite{MQ86}.

\begin{acknowledgement}It is our pleasure to thank Christian B\"ar and Steve Rosenberg for
helpful comments on the draft of this paper. 
\end{acknowledgement}
\section{Preliminaries and Notation} 
\label{sec:prelim}

We review the key facts needed from Riemannian geometry and fix
notation, most of which follows Berline, Getzler and Vergne
\cite{BGV04}.

\para{Notation for Riemannian geometry}{notation-rg}
Let  $M$ be a compact oriented $2m$-dimensional Riemann manifold.
In a coordinate patch let
$ \partial_\mu$ be the corresponding basis of tangent fields,  $\dx^\mu$
be the dual basis of  one-forms\footnote{The element of the dual basis
  is more commonly denoted $dx^\mu$. We use $\dx^\mu$ in anticipation
  of the  interpretation in terms of supersymmetric variables
  in~\ref{par:susy} below.}, and  $\oppartialx{\mu}$ be the
odd derivation on forms defined by
\[\oppartialx{\mu} \dx^\nu = \delta_\mu^\nu.\]

The metric $g_{\mu \nu}= (\partial_\mu, \partial_\nu)$ determines
Christoffel symbols 
\be
\Gamma_{\mu \nu}^\gamma=\frac{1}{2}g^{\gamma \eta}(\partial_\nu g_{\mu \eta} + \partial_\mu
g_{\nu \eta} - \partial_{\eta} g_{\mu \nu})=\frac{1}{2}g^{\gamma \eta}(g_{\mu
\eta,\nu} +  
g_{\nu \eta,\mu} -  g_{\mu \nu ,\eta}),
\ee{gamma-def}\noindent
(indices after the comma denote
differentiation in that coordinate) in terms of which  the Levi-Civita connection is
\[\Nabla_\mu(Y^\nu \partial_\nu)= (\partial Y^\nu/\partial
x^\mu) \partial_\nu + \Gamma_{\mu \eta}^\nu Y^\eta \partial_\nu.\]

The operator $\Nabla$ extends to a one-form with values in
differential operators
on forms by 
\[\Nabla_{\mu} = \iota_{\mu}d - \Gamma_{\mu \nu}^\eta
\dx^{\nu} \oppartialx{\eta}.\]
The curvature $R$ of the Levi-Civita connection is a smooth two-form with values
in linear transformations  on the fiber. Acting on the coordinate basis, it is
\[  R(\partial_\pi, \partial_\eta) \cdot \partial_\mu= R_{\pi \eta
  \mu}^{\triplegap \nu}\partial_\nu,\]
where
\be
R_{\mu \nu \gamma}^{\triplegap \delta}=
\Gamma_{\nu \gamma ,\mu}^\delta - \Gamma_{\mu \gamma ,\nu}^\delta +
\Gamma_{\mu \chi}^\delta 
\Gamma_{\nu \gamma}^\chi - \Gamma_{\nu \chi}^\delta \Gamma_{\mu \gamma}^\chi.
\ee{R-def-coords}\noindent

We will freely raise and lower all four indices on $R$ with the
metric, keeping track of the order by spacing.  With this
convention, 
the symmetries of $R$ are
\[
R_{\mu \nu \pi \eta} = R_{\pi \eta \mu \nu} = -R_{\nu \mu \pi \eta}
 \qquad 
R_{\mu \nu \pi}^{\triplegap \eta}  + R_{\pi \mu \nu }^{\triplegap
  \eta}  + R_{\nu \pi \mu }^{\triplegap \eta}  =0. 
\]

The Ricci tensor  is
\be
\Ricci_{\sigma \tau}= R_{\sigma \mu \tau}^{\triplegap \mu}.
\ee{Ricci-def}\noindent

\para{Laplace-Beltrami and heat kernels}{gen-hk}

The Laplace-Beltrami operator $\Delta$ on the space $\Omega(M)$ of forms
is 
\be
\Delta= - g^{\mu \nu}(\Nabla_\mu \Nabla_\nu- \Gamma_{\mu \nu}^\sigma
\Nabla_\sigma)
-\Ricci_\eta^\pi \dx^\eta \oppartialx{\pi}
-\frac{1}{2} R_{\mu \eta}^{\doublegap \nu \pi} \dx^\mu \dx^\eta
\oppartialx{\nu} \oppartialx{\pi}.
\ee{lb-def}

The evolution operator $e^{-t\Delta/2}$ is a semigroup of operators on
$\Omega(M)$ depending on a parameter $t \in [0,\infty)$ such that
for  $\alpha\in \Omega(M)$ $\alpha_t=e^{-t\Delta/2} \alpha$ is a
solution to the heat equation 
\[(\Delta/2 + \partial_t)\alpha_t=0\]  
with $\alpha_0=\alpha$ as initial conditions. 

The \emph{heat kernel,} a smooth map $K_\Delta$
from $(0,\infty)$ to sections of $\Omega(M \times M)$, provides an
integral representation of the time evolution operator. Explicitly, for
  $\alpha \in \Omega(M)$ 
\[(e^{-t\Delta/2}\alpha)(x)=K_\Delta * \alpha = \int_{y \in M}
K_\Delta(x,y;t) \alpha(y)\] 
where on the right-hand side we wedge the forms over $y$ together,
take the top form 
piece, and integrate over the second factor of $M.$  In general
operators on $\Omega(M)$ are represented by forms in
$\Omega(M\times M),$ with operator composition    
\be 
K_1 * K_2(x,z)= \int_{y \in M}
K_1(x,y) K_2(y,z).  
\ee{composition-def}

\para{Riemann Normal Coordinates}{rn-bundle}

Orthonormal coordinates on $T_yM$ extend via $\exp_y$ to coordinates
on a patch of $M$ called Riemann normal coordinates.  In Riemann normal
coordinates lines from the origin are geodesics with length consistent
with the coordinates, and the following
hold, where $\vec{x}=x^\mu \partial_\mu$ is the tangent
vector at $y$ corresponding to $x$ and $|\vec{x}|$ is its length
\be
g^{\mu \nu} (x) = \delta^{\mu \nu} - \frac{1}{3} R_{\gap \sigma \gap
  \tau}^{\mu \gap \nu}(0) x^\sigma x^\tau 
+ O(|\vec{x}|^3),
\ee{metric-2-order}
\be
\Gamma_{\mu\gamma}^\delta(x) = -\frac{1}{3}\left[R_{\mu \nu \gamma}^{\triplegap
  \delta}(0) 
+ R_{\gamma \nu \mu}^{\triplegap \delta}(0)\right] x^\nu + O(|\vec{x}|^2).
\ee{gamma-1-order}
Finally,  any vector $v \in T_xM$ defines two vectors in $T_y M$: the
first is $\vec{v} = ( d \, \exp_y)^{-1} v$; the second is the parallel
translate $\pt{v}$ of $v$ along the geodesic from $x$ to $y$. These
are related by
\be
\pt{v}= \vec{v}   + \frac{1}{6} R(\vec{x}, \vec{v})\cdot
\vec{x} + O(|\vec{x}|^3)|v|.
\ee{pt-2-order}

Whenever we work in Riemann normal coordinates we  implicitly
restrict attention  to a patch within the
injectivity radius of the center, small enough that there is a
unique  geodesic from the center to each point.

\para{Supersymmetric variables}{susy}

If $V$ is a vector space, we  represent elements of $\Lambda(V^*)$
by formulas involving an anticommuting element $\dx$ of
$V.$  For example, given  a basis $e_1, \ldots , e_n$  of $V,$ an antisymmetric matrix
$\omega_{\mu \nu}$ determines an element $\omega(\psi)$ of $\Lambda^2(V^*)$ via
$\omega(\psi) = \frac{1}{2} \omega_{\mu \nu}\dx^\mu  \dx^\nu,$ with
 $\dx=\dx^\mu
e_\mu$. 
In the latter expansion of $\dx$,  each $\dx^\mu$ is an
anticommuting numerical variable. On the other hand, each $\dx^\mu$ in
the expansion of an element of $\Lambda(V^*)$ is
 a map sending the real element $v \in V$ to a real number; namely,
 its component $v^\mu$ in the given basis. Thus
$\dx^1, \ldots, \dx^n$ also represents the basis of $V^*$ dual to $e_1,
\ldots , e_n.$ In this interpretation $\dx=\dx^\mu e_\mu$ is then an
expression for the identity map $dx^\mu e_\mu$ on $V$ composed with
the natural map from $V$ to the exterior algebra $\Lambda(V)$. In calculations
it is usually easier to work with $\dx$ as denoting an anticommuting
tangent vector; to interpret the resulting expressions, it is helpful
to remember it means this identity map.

By the same token we can consider $\rho$ as an anticommuting variable
in $V^*$ which will be used in formulas representing elements of
$\Lambda(V).$ In this context $\rho_\mu$ replaces $e_\mu$ in the usual
expressions. Equivalently, $\rho$ represents  the identity map on $V^*.$

Most often $\rho$ will be used inside a Berezin
integral. This integration is defined for $f$ an anticommuting
polynomial,  in terms of a volume form on $V$, by $\bint
f(\rho)$ is the volume of the $\dim(V)$ degree piece of $f.$  For
example if $V$ is $2m$ dimensional, with a basis  chosen so $\dx^1
\cdots \dx^{2m}$ is the volume form, and $g(\dx)$ is an anticommuting
polynomial in $\dx$, then the Berezin integral over $\rho$ is
\begin{eqnarray*} \bint e^{i\bracket{\rho, \dx}}g(\dx) \, d\rho &=& \bint \sum_k \frac{\left(i
  \rho_\mu \dx^\mu\right)^k}{k!} \, \sum g_{\nu_1 \cdots \nu_l} \dx^{\nu_1}
\cdots \dx^{\nu_l} \, d\rho\\
&=& \bint \frac{(-1)^m}{(2m)!} \,  \rho_{\mu_1} \dx^{\mu_1} \cdots
\rho_{\mu_{2m}}\dx^{\mu_{2m}} \, \sum g_{\nu_1 \cdots \nu_l} \dx^{\nu_1}
\cdots \dx^{\nu_l} \, d\rho\\
&=& \dx^1 \cdots \dx^{2m} \, \sum g_{\nu_1 \cdots \nu_l} \dx^{\nu_1}
\cdots \dx^{\nu_l} \\
&=&g(0) \dx^1 \cdots \dx^{2m}.
\end{eqnarray*}
The right-hand side denotes the $0$-degree part of $g$ times the volume
form on $V.$ 

In this paper $\dx$ and $\rho$ will be anticommuting elements of the tangent and cotangent spaces, respectively, at
a point in $M,$ so that the formulas involving them will describe forms on $M.$ Two examples serve to illustrate Berezin integration in this context and to provide formulas we will require in Section~3. With $\dx_x$, $\rho^y$ and $\dx_y$ denoting anticommuting tangent and cotangent vectors at points $x$ and $y$ in $M$, $\pt{\dx}_x$ representing the parallel transport of $\dx_x$  from 
$x$ to $y$ along some path connecting them, and
$\alpha \in \Lambda(T^*_yM)$,
\be\bint \! \bint e^{i\bracket{\rho^y, \pt{\dx}_x -\dx_y}}\alpha \, d\rho^y d\dx_y= \, \pt{\alpha}.
\ee{bint-pt}
Here $\pt{\alpha}$ is $\alpha$
parallel transported along the given path.    
Thus we have an
operator that can implement parallel transport. (Of course, parallel
transport could be replaced by any linear map.) The key to this
calculation is that the coefficient of the top form in $\rho$ is
proportional to $[(\pt{\dx}_x)^{1} -\dx_y^{1}] \cdots
[(\pt{\dx}_x)^{2m} -\dx_y^{2m}]$. The top-form piece of the
product of this with $\alpha$ will include terms like
$(\pt{\dx}_x)^{1}\dx_y^{2}\dx_y^{3} (\pt{\dx}_x)^{4}\cdots
\dx_y^{2m}\alpha_{1 4}(y) \dx_y^1 \dx_y^4 $ which will contribute the
term $\dx^1_y \cdots
\dx^{2m}_y \alpha_{1 4}(y)(\pt{\dx}_x)^{1}(\pt{\dx}_x)^{4} $ leading,
after integration with respect to $\dx_y$,
to   $\pt{\alpha}$ on the right-hand side.
Likewise, for $\mu
\in \{1, \ldots , 2m\},$

\be\bint \! \bint  \rho^y_\mu e^{i\bracket{\rho^y, \pt{\dx}_x -\dx_y}}\alpha \, d\rho^y d\dx_y=i \, 
\pt{\left(\oppartialx{\mu}\alpha\right)} . 
\ee{bint-iota}

In this notation, $f(x,\dx_x)$, for $f$ smoothly varying in $x$ and an
antisymmetric multinomial in $\dx_x$, corresponds to a smooth
differential form $f$ on $M$. Moreover,
$\int \bint f(x,\dx_x) \, d\dx_x dx$ is the integral $\int_M f$ of the top-form
part of $f$ over $M$. 
\section{Discrete Approximation to the SUSYQM Lagrangian}

 In this section we define  a sequence of finite-dimensional
subspaces of the space of paths in $M$ 
on which we interpret the $N=1$ supersymmetric quantum mechanical
Lagrangian as a form. This form describes a kernel which is
an operator product of a number of copies of a simpler kernel described by a form $\MQ$
on a $4m$-dimensional space. We will ultimately apply B\"ar \& Pf\"affle's arguments to show
that, as the dimension of the subspaces increases, the product of kernels
converges uniformly to the kernel of the  Laplace-Beltrami
heat operator.
\para{Short geodesics}{short-gdsc}

A \emph{short geodesic} is a geodesic of length less than the
injectivity radius of $M.$ The space of short geodesics is isomorphic
to $M^{(2)},$ the subspace of $M^2$ consisting of pairs of points
within the injectivity radius of each other. (We  take our paths as
oriented but not parameterized; later we will choose
 parameterizations).  Let $\Path_n$ denote the space of $n$-segment piecewise
short geodesic paths in $M$,  and let $\Path_n(x,y)$ denote the subspace of those
going from $y$ to $x.$ $\Path_n$ is isomorphic to $M^{(n+1)},$ the subspace of
$M^{n+1}$ in which each successive point is within the injectivity
radius of the previous.  

If $\sigma_t$ is a short geodesic in $\Path_1$ the isomorphism
with $M^{(2)}$ sends $\sigma$ to $(x,y)$, where $x = \sigma_1$ and $y=
\sigma_0$. Note the 
unconventional choice of a path going from $y$ to $x.$ This is necessitated by
the standard conventions of kernels and operators. 

\para{Tangents to short geodesics}{tngnt-gsdc}
If  $\sigma_t$ is a geodesic, represent a tangent vector to it in the
space of geodesics 
by a tangent field $\dx_t \in T_{\sigma_t}M$ along $\sigma.$\footnote{In what
follows, the components of this vector field could be either real or
anticommuting numbers. Since our application of the lemma
below will be to the anticommuting case, we use $\psi$ to denote a
generic vector.} Let
$\pt{\dx}_t \in T_{\sigma_0}M$ be the parallel translate of $\dx_t$ from
$\sigma_t$ to $\sigma_0$ along $\sigma$ according to the Levi-Civita
connection.  
Suppose $\sigma$ from $t=0$ to $t=1$ is a short geodesic, and take
$\dx_t$ to be tangent to a one-parameter family of short geodesics.
Since each geodesic in this family is determined by its
endpoints, $\dx_t$ should be determined by $\dx_0$ and $\dx_1.$ In fact,
\begin{lemma}\label{lm-gdsc-tngt}  If $\sigma_t$ is a geodesic path mapping $[0,1]$ to $M,$ $\dx_t$
is a tangent field along $\sigma,$ $d=d(\sigma_0, \sigma_1)$ is the distance
between the endpoints of $\sigma$, and $|\dx|=\max(|\dx_0|, |\dx_1|)$, then
\eqa{gdsc-tngt}{
\pt{\dx}_t &= t\pt{\dx}_1 + (1-t) \dx_0 + \frac{t^3-t}{6}
R(\dot{\sigma}_0,\pt{\dx}_1)\cdot \dot{\sigma}_0\nonumber \\
& \qquad  -
\frac{t^3-3t^2+2t}{6}R(\dot{\sigma}_0,\dx_0)\cdot \dot{\sigma}_0 +
O(d^3)|\dx|
}
where $R$ is computed at $\sigma_0,$  and $\dot{\sigma}_t= \partial_t
\sigma_t.$ 
\end{lemma}

\begin{proof}{} 
Since the result is linear in $\dx_1$ and $\dx_0,$ we  prove it
when $\dx_0=0.$  The case $\dx_1=0$ and thus the general case follow
from reversing the parameterization.

In Riemann normal coordinates centered at $\sigma_0,$  $\psi_1 = (d \,
\exp_{\sigma_0}) \vec{\dx}_1$ for some $\vec{\dx}_1 \in T_{\sigma_0}M$. 

Extend $\vec{\dx}_1$ to a path of tangent vectors as
\[\vec{\dx}_t= t \vec{\dx}_1.\]
Note that because lines through the origin are geodesics in Riemann
normal coordinates, this path of tangent vectors describes a tangent
vector to the space of geodesics. 

Applying Eq.~\eqref{eq:pt-2-order} to $\vec{\dx}_t$ and $\vec{\dx}_1$
gives
\[\pt{\dx}_t= \vec{\dx}_t + \frac{t^2}{6} R(\dot{\sigma}_0,
\vec{\dx}_t) \cdot \dot{\sigma}_0 + O(d^3)|\dx| \]
\[\pt{\dx}_1= \vec{\dx}_1 + \frac{1}{6} R(\dot{\sigma}_0,
\vec{\dx}_1) \cdot \dot{\sigma}_0 + O(d^3)|\dx| \]
so
\[\pt{\dx}_t= t \pt{\dx}_1 +
\frac{t^3-t}{6}R(\dot{\sigma}_0,\pt{\dx}_1)\cdot
\dot{\sigma}_0 + O(d^3)|\dx|.\]

Reversing the parameterization and assuming $\dx_1 = 0$ introduces the
terms $(1-t) \dx_0 + \frac{(1- t)^3-(1-t)}{6}R(\dot{\sigma}_1,\pt{\dx}_0)\cdot
\dot{\sigma}_1 $, where the parallel transport is from $\sigma_{0}$
to $\sigma_{1}$. After parallel transporting back to $\sigma_0$ in the
second term, these become the additional terms the lemma requires. Note
this substitution is permitted to the given order, since $\dot{\sigma}$ is parallel
along $\sigma$, and the difference between applying the curvature and
metric at
$\sigma_1$ and applying them at $\sigma_0$ after parallel transport is
of order $d^3|\psi|$. 
\end{proof}
\begin{remark}
The scale of the parameterization is of course arbitrary in the above
lemma. $\dx$ is determined by its value at any two points of $\sigma$, and
Eq.~\eqref{eq:gdsc-tngt} continues to describe this dependence with the parameter
$t$ adjusted appropriately.
\end{remark}

\para{The SUSYQM Lagrangian}{susy-form}
The  action  for $N=1$ supersymmetric quantum mechanics on the
manifold $M$ is
\[S(\sigma, \dx, \rho, t) = \int_0^t \left(-\frac{\dot{\sigma}_r^2}{2} + i \bracket{\rho^r,
    \left(\Nabla_{\dot{\sigma}}\dx\right)_r} -\frac{1}{4} \left(\rho^r,R(\dx_r,
    \dx_r)\cdot \rho^r\right) \right) dr.\]
where $\sigma$ is an element of the space of paths in $M$, $\psi_r$ is
an anticommuting element of the tangent to the space of paths, and $\rho^r$ is
an anticommuting  variable modeled on the
dual to the tangent space of the space of paths. In a  pairing $\int_0^t\bracket{\rho^r,
  \psi_r} dr$, 
the end result
is (at least formally)  a one-form on the space of paths with values
in linear functions in $\rho$. The partition function for SUSYQM
on $M$ is
\[
Z = \int \bint e^{S(\sigma, \dx, \rho, t)} .
\]
The (formal) Berezin integration in $\rho$ 
produces a form on the space
of paths. 
The ``top form
piece''  of this form is integrated over the space of
paths to give the partition function. Taking the paths to have fixed
endpoints, the partition function is a path integral representation
for the kernel of  the time evolution
operator or the Feynman propagator.

Given a family of paths $\sigma$, we may think of $\dot{\sigma}$ and
  $\dx$ as vector fields on $M$, which must necessarily commute, since
  the paths locally define coordinate curves which are integral curves
  for $\dx$ and $\dot{\sigma}$. Thus, in the action  we  may replace
$\Nabla_{\dot{\sigma}}\dx$ with $\Nabla_{\dx}\dot{\sigma}.$  We thereby
recognize the Lagrangian as (formally) exactly the Mathai-Quillen Thom form on the tangent
bundle to the space of paths, pulled back by the
section $\dot{\sigma}$. 
The connection is the Levi-Civita connection
determined by the metric $\int_0^t (X_t, Y_t) dt $.  This observation
and its formal consequences are due to Blau~\cite{Blau93}.

It is of course the integral over the infinite-dimensional space of
paths that makes the links between the heat kernel, the partition
function, and a Mathai-Quillen integral purely formal. 
However, if we 
interpret the path integral by restricting it to a sequence of finite-dimensional subspaces that in a
reasonable sense approximate the whole space of paths, the arguments
are correct on the finite-dimensional approximating spaces.  

We approximate the space of continuous paths $\sigma: [0,t]
\rightarrow M$ with $\sigma(0) = y$ and $\sigma(t) = x$ by $\Path_n(x,y).$  We choose positive numbers $t_1,
\ldots, t_n$ such that $t=\sum_{i=1}^n t_i$ and parameterize each path
in $\Path_n(x,y)$  so that the first segment is the image of $[0,
t_1]$ parameterized 
proportionally to arclength (so the segment is a paramaterized geodesic),
the second segment is the image of $[t_1,t_1+t_2]$ parameterized proportionally to
arclength, and so forth. Let $\Path_n(x,y; t_1, \ldots, t_n)$ denote
the space of paths in $\Path_n$ parameterized in this way so that the
parameter length of the $i$th geodesic segment is $t_i$.  In the computation of the approximation to the partition
function, $\dx$ will become an anticommuting vector tangent to
$\Path_n$. 
This tangent space has dimension $2m(n+1)$, since it consists of
 vectors $\dx_1, \ldots , \dx_{n+1}$ with $\dx_{i} \in T_{x_i}M$
and with the $x_i$ denoting the $(n+1)$ endpoints of the
geodesic segments. 

The
situation with $\rho$ is a bit more complicated:  The quantum mechanical
state space consists of sums of anticommuting
polynomials in $\dx_i^\mu$ with coefficients depending on $x_i$; these
correspond to forms on $M^{n+1}$. 
 The
path integral will give a kernel of the 
time evolution operator which will act  on the form on $\sigma_0$
representing the initial state. 
 Thus we should think of the form at
$\sigma_0$ as already being determined, so that the space in which $\dx$
lives is the space of all tangent vectors extending a given tangent
vector at $\sigma_0.$   The variable $\rho$ should thus not be a dual tangent
vector at each of $n+1$ terminal points of the geodesic pieces, as we
might naively expect, because it should have no component dual to the tangent
space at $\sigma_0$.  Thus $\rho$ will
consist of  dual
vectors $\rho^1, \rho^2, \ldots, \rho^n,$ with each $\rho^i$ an
anticommuting element in
$T_{x_i}^*M,$
 where $x_i$ is the final point of the $i$th
  segment.  The pairing of $\rho$ and $\psi$ is given by
\[\int_0^t \bracket{\rho^r, \dx_r} dr= \sum_{i=1}^n t_i \bracket{\rho^i,
   \dx_{\xsi}}.\]
Note that in local coordinates $x_i^\mu$ in a neighborhood of $x_i$,
the pointwise pairing on the right-hand side is $\bracket{\rho^i,
   \dx_{\xsi}} = \rho_\mu^i \dx_{\xsi}^\mu$.

Thus the natural restriction of the path integral to the space of
piecewise geodesic paths is 
\eqa{susy-n-def}{
&\MQ_n(x,y;t_1, \ldots, t_n))=\prod_{i=1}^n\left(2 \pi t_i
\right)^{-m}\int_{\Path_n(x,y;t_1, \ldots, t_n)}
\bint \cdots \bint  \nonumber \\
& \qquad  \qquad\exp\Bigg[\sum_{i=1}^n 
  -\frac{|\dot{\sigma}_{i}|^2}{2} t_i   +  i \, t_i  \bracket{\rho^i,
  \left(\Nabla_{\dot{\sigma}} \dx
      \right)_{\xsi}} \nonumber \\
&\qquad \qquad  \qquad  - \frac{t_i}{4}\left(\rho^i,
   R(\dx_{\xsi}, \dx_{\xsi} )  \cdot \rho^i
\right)  \Bigg] \, d\rho^1 \cdots d\rho^n
}
where $\dot{\sigma}_{i}$ denotes tangent  the final point $x_i$ of the
$i$th geodesic segment.

The normalization
factor out front is chosen to make the trivial case $M = \RR^{2m}$ with
the Euclidean metric work
out right.

 The  time evolution operator associated to the kernel above is a composition of
operators, each corresponding to one geodesic piece and each having
$\MQ_1$ as its kernel.  In other words (suppressing the spatial variables)
\[\MQ_n(t_1, t_2, \ldots, t_n)= \MQ_1(t_1) * \MQ_1(t_2) * \cdots
*\MQ_1(t_n).\]

\para{Expressing $\MQ_1$ as a form on $M^{(2)}$}{susy-on-m2} 
In this section we explicitly evaluate the form $\MQ_1$, in terms of
geometric invariants.
It is natural to rescale the parameterization length to $1$, and adjust
the meaning of
$\dot{\sigma}$ accordingly, to obtain the 
following form on the same path parameterized from $0$ to $1$
\begin{align*}\MQ(x,y;t)&= \left(2 \pi t \right)^{-m}
\bint  \exp\Big[
  -\frac{|\dot{\sigma}|^2}{2t}  \nonumber \\
& \qquad  + i \bracket{\rho^x,
   \left(\Nabla_{\dot{\sigma}}\dx\right)_x}  - \frac{t}{4}\left(\rho^x,
   R(\dx_x, \dx_x ) \cdot \rho^x
\right) \Big] \, d\rho^x. 
\end{align*}
 This is a  form on  $\Path_1\iso
M^{(2)},$ and can be expressed as such. First $\dot{\sigma}$   in Riemann normal
coordinates centered at $x$ is $-\vec{y}.$  
From Lemma~\ref{lm-gdsc-tngt}, by taking the derivative with respect to $t$ at $t=1$ in Eq. 2.1 and parallel transporting everything to $x=\sigma_1,$ we get
\[\left(\Nabla_{\dot{\sigma}}\dx\right)_x = \dx_x - \pt{\dx}_y
+\frac{1}{3} R(\vec{y}, \dx_x) \cdot\vec{y}  +
\frac{1}{6} R(\vec{y}, \pt{\dx_y}) \cdot \vec{y}
+ O(|\vec{y}|^3)|\dx|.\] 
So
\begin{align*}
&\MQ(x,y;t) = \left(2\pi t\right)^{-m} \bint \exp \left[
-\frac{|\vec{y}|^2}{2t}   -\frac{t}{4}
\left(\rho^x,  
R({\dx}_x,{\dx}_x) \cdot \rho^x \right)  \right.
\\
&\qquad  \left. i \bracket{\rho^x, \dx_x -
  \pt{\dx_y} + \frac{1}{3} R(\vec{y}, \dx_x) \cdot\vec{y}  +
\frac{1}{6} R(\vec{y}, \pt{\dx_y}) \cdot \vec{y}} +
O(|\vec{y}|^3)\right]\, d\rho^x.
\end{align*}

\para{Shifting $\dx_y$ to $\dx_x$}{shift-dy-dx}
Suppose $\eta$ and $\pi$ are indices for an orthonormal basis of
$T_yM.$  Suppose $f(\rho)$ is an anticommuting polynomial in the $\rho_1,
\ldots, \rho_{2m} $ excluding $\rho_\eta,$ and $g(\dx)$ is an
anticommuting polynomial in the $\dx^1, \ldots , \dx^{2m}$ excluding
$\dx^\pi.$  Then 
\eqa{}{
& \bint i \rho_\pi \left( \dx_x -
  {\pt{\dx}_y}\right)^\eta f(\rho) g(\dx) \exp\left[ i
  \bracket{\rho,\dx_x - \pt{\dx}_y}\right] \, d\rho = \nonumber \\
& \qquad \bint  f(\rho)
g(\dx)  \delta_\pi^\eta
\exp\left[ i 
  \bracket{\rho,\dx_x - \pt{\dx}_y}\right] \, d\rho. \nonumber} 
   In
 particular,  within an integral against $ \exp\left[ i
  \bracket{\rho,\dx_x - \pt{\dx}_y}\right]$,
\[\frac{i}{6} \bracket{\rho,R(\vec{y}, \pt{\dx_y}) \cdot
  \vec{y}} =
\frac{i}{6} \bracket{\rho,R(\vec{y}, \dx_x) \cdot \vec{y}}
-\frac{1}{6} (\vec{y}, \Ricci\cdot \vec{y}).
\]
So defining 
\be
H(x,y;t)=(2\pi t)^{-m} \exp\left[-\frac{1}{2t} d(x,y)^2 \right]\
\ee{h-def}
for $x,y \in M$ within the injectivity radius of each other and $t>0$
(and zero otherwise)  this gives
\eqa{mq-global}{
\MQ(x,y;t) &= H(x,y;t) \bint \exp \Big[ i \bracket{\rho^x, \dx_x -
  \pt{\dx}_y  + \frac{1}{2} R(\vec{y}, \dx_x) \cdot\vec{y}
}\nonumber \\ 
&\qquad -\frac{1}{6} (\vec{y}, \Ricci\cdot \vec{y})   -\frac{t}{4} \left(\rho^x, 
R(\dx_x,\dx_x) \cdot \rho^x \right) + O(|\vec{y}|^3)\Big] \, d\rho^x.
}

\para{Mathai-Quillen on paths and loops}{mq-loops}
The vector bundle $TM \times M \to M \times M=M^2$ restricts to a bundle over
the open submanifold $M^{(2)} \iso \Path_1(M).$  A natural section of
this bundle assigns to each $(x,y) \in M^{(2)}$ the tangent vector
$\vec{y}$ at $x$ (or equivalently, in terms of $\Path_1(M),$ the
vector $-\dot{\sigma}_1$).  The Levi-Civita connection on $M$ extends
to a connection on this bundle, in terms of which one can easily
verify that
$\MQ_1(t)$ gives the pullback of the Mathai-Quillen Thom form on this
bundle via the section.  We note that for finite positive $t$ this gives a
closed form on $M^{(2)},$ but not a compactly
supported closed form.
Likewise the form on $\Path_{n}(M) \iso
M^{(n+1)}$ whose integral gives $\MQ_n$ is the pullback by the
corresponding section of the
Mathai-Quillen form on the bundle $T(M \times M \times \cdots \times M) \times M
\to M^{n+1}$ restricted to the subset $M^{(n+1)}$  (after
absorbing the $t_i$'s into the metric on the various factors of $M$).

Instead of paths we can consider piecewise geodesic \emph{loops.}
Here it is natural to consider the kernel Eq.~\eqref{eq:susy-n-def}
with not only the points $x_0$ and $x_n$ identified but also $\dx_0$
and $\dx_n$ identified. That is, we identify $x$ and $y$ and wedge the
form over $x$ with the form over $y$ (the form over $x$ coming
first). The integral of the resulting form on $M$ is the supertrace of
the kernel on the left-hand side of Eq.~\eqref{eq:susy-n-def}. Proving
that as $n$ goes to infinity the latter kernel converges to the heat
kernel will show this integral converges to the supertrace of the heat
kernel.  The ability to connect the supertrace of the heat kernel to
the integral of the pullback of the Mathai-Quillen form for a tangent
bundle, through an intervening limit, is strong circumstantial
evidence that this is a productive way of interpreting the
supersymmetric path integral.

\section{Strong Convergence of the Time Evolution Operator}

B\"ar and Pf\"affle \cite{BP07} offer a rigorous expression for
various heat kernels as a kind of path integral.  Specifically they
use  a form of Chernoff's theorem to 
prove the following result:

\begin{theorem}[B\"ar, Pf\"affle]\label{th-op-cnvg} Suppose $K(x,y;t) \in
E_x \tensor E_y^*$ is a smooth
one-parameter family of kernels (with positive real parameter $t$)
representing the family of operators ${\mathfrak{K}}(t)$ on a Euclidean vector
bundle $E$  that satisfy the
following three assumptions:
\begin{enumerate}
\item $||{\mathfrak{K}}(t)||=1+ O(t)$ for small $t,$ where the norm is as
  an operator on the space of smooth functions with the supremum norm.
\item On each  $\alpha \in \Gamma(M,E)$ 
\[\lim_{t\to 0} ({\mathfrak{K}}(t)\alpha -\alpha)/t \to -\frac{\Delta}{2} \alpha,\]
in the supremum norm where $\Delta$ is a generalized Laplacian on $E.$
\item For each $y$
\[
 \lim_{t \to 0} K(x,y;t)= \delta(x,y)
\]
 as  a distribution.
\end{enumerate}
If $t_1, t_2, \ldots,
t_n$ is called a partition, then for any sequence of partitions in
which $\max_i t_i \to 0$ and $\sum_i t_i \to t$ and for any form
$\alpha$ on $M$
\[\lim {\mathfrak{K}}(t_1) {\mathfrak{K}}(t_2) \cdots {\mathfrak{K}}(t_n) \alpha = e^{-t\Delta/2}
\alpha.\]
Moreover, for some such sequence of partitions 
\[\lim K(t_1) * K(t_2) * \cdots * K(t_n) \to K_\Delta(x,y;t)\]
uniformly, where $K_\Delta$ is the heat kernel of $\Delta,$ i.e. the
kernel of $e^{-t\Delta/2},$  and we
suppress the spatial variables in $K.$
\end{theorem}

\begin{remark}\label{rm-half-delta}  
B\"ar and Pf\"affle work with $\Delta$ rather than $\Delta/2,$ which
of course amounts to nothing more than a rescaling of $t$ by a factor
of $2.$  However, in the usual scaling of the physics literature, the
time evolution operator corresponds to $e^{-t\Delta/2},$ so we follow
this convention.
\end{remark}
 \para{Applying the theorem to $\MQ$}{BP-MQ}
B\"ar and Pf\"affle apply this theorem to operators constructed from
heat kernel asymptotics to give their path integral formulation. It is
possible to relate  Eq.~\eqref{eq:mq-global}  to
the kernel in their Theorem 6.1 (note that their paths are
parameterized in the opposite direction, and thus signs on all
integrals are reversed), thus showing that supersymmetric quantum
mechanics path integral restricted to piecewise short geodesic paths
approaches the heat kernel for the Laplace Beltrami operator on
forms as the number of pieces goes to infinity (for certain sequences
of parameterization lengths).  Instead we will  check directly that the SUSYQM
Lagrangian satisfies the assumptions of Theorem~\ref{th-op-cnvg},
thus achieving the same result.  The check
is a simple
calculation that involves no sophisticated understanding of heat
kernel asymptotics and seems closer in spirit to path integral
arguments.

Write $\MQ(t)$ for $\MQ(x,y;t)$ when the spatial variables are to be
understood, and $\MQO(t)$ for the operator represented by this kernel.

\begin{proof}{ of Assumption 1}\label{par:pf-op-norm} The operator norm of
$\MQ(t)$ is $1+O(t).$
 By compactness we can check this pointwise at each $x,$ and because
 $\MQ(t)$ is zero outside the injectivity radius we can do the
 calculation inside a coordinate patch in Riemann normal coordinates.
 It suffices to let $\MQO$ act on a function times a covariantly constant form, and
 the result follows from the fact that $H(x,y;t)$ has operator norm
 $1.$ 
\end{proof}

\begin{proof}{ of Assumptions 2 and 3} \label{par:pf-mq-1}  If $\alpha$
  is a form on $M,$ we
 must show
\[\lim_{t\to 0} \left(\MQO(t) \alpha -\alpha\right)/t= -\frac{\Delta}{2} \alpha\]
where $\Delta$ is the Laplace-Beltrami operator on forms Eq.~\eqref{eq:lb-def}.
Again, we may check at a specific point $x,$ and we may assume $\alpha$ is
zero outside the geodesic neighborhood of $x.$ We may also assume
$\alpha$ is simply a function times a covariantly constant form, so that $\pt{\alpha}(y,\dx_y) =
f(y)\alpha(x,\dx_x) ,$ where the parallel transport from $y$ to $x$ is along the minimal geodesic.

Working in Riemann normal coordinates centered at
$x$ so that
\be{\det}^{1/2}(g)(\vec{y})= 1 + \frac{1}{6} \Ricci_{\sigma \tau} y^\sigma y^\tau
+ O(|\vec{y}|^3),\ee{rndet}
and writing $H(\vec{y};t)$ for the expression of $H(x,y;t)$
in these coordinates,
gives
\begin{align*}
\MQO(t) \alpha &= \int H(\vec{y};t) \bint \! \bint \exp\left[ i \bracket{\rho^x, \dx_x -
  \pt{\dx}_y + \frac{1}{2} R(\vec{y}, \dx_x) \cdot\vec{y}}\right.\\
&\qquad \left.-\frac{1}{6} (\vec{y}, \Ricci\cdot \vec{y})  -\frac{t}{4} \left(\rho^x, 
R(\dx_x,\dx_x) \cdot \rho^x \right) + O(|\vec{y}|^3)\right] \\
&\qquad \cdot \alpha(\vec{y},\dx_y) \, d\rho^x d\dx_y d\vec{y} \\ 
&= \int H(\vec{y};t) \bint \! \bint \exp\left[ i \bracket{\rho^y, \pt{\dx}_x -
  \dx_y + \frac{1}{2} R(\vec{y}, \pt{\dx}_x) \cdot\vec{y}}\right.\\
&\qquad \left.-\frac{1}{6} (\vec{y}, \Ricci\cdot \vec{y})  -\frac{t}{4} \left(\rho^y, 
R(\pt{\dx}_x,\pt{\dx}_x) \cdot \rho^y \right) + O(|\vec{y}|^3)\right]\\
&\qquad \cdot \alpha(\vec{y},\dx_y) \, d\rho^y d\dx_y d\vec{y}\\ 
&= \int H(\vec{y};t) \bint \! \bint \left[1 - \frac{1}{6} \Ricci_{\sigma \tau}
  y^\sigma y^\tau + \frac{i}{2}\rho^y_\tau R_{\pi \eta
    \sigma}^{\triplegap \tau}y^\pi (\pt{\dx})_x^\eta y^\sigma 
+ O(|\vec{y}|^3)\right] \\
&\qquad\cdot \left( 1 - \frac{t}{4}\rho_\nu^y  
R_{\mu \eta}^{\doublegap \nu
  \pi}(\pt{\dx})^\mu_x(\pt{\dx})^\eta_x 
\rho^y_{\pi}  + O(t^{3/2})\right)  \exp\left[ i \bracket{\rho^y,
  \pt{\dx}_x - \dx_y}\right]  \\
&\qquad \cdot \alpha(\vec{y}, \dx_y) \,
d\rho^y d\dx_y d\vec{y} \\
&= \int H(\vec{y};t) f(y) \left[1 - \frac{1}{6} \Ricci_{\sigma \tau}
  y^\sigma y^\tau + \frac{1}{2} R_{\pi \eta
    \sigma}^{\triplegap \tau}y^\pi \dx_x^\eta y^\sigma \oppartialx{\tau} 
+ O(|\vec{y}|^3)\right] \\ 
&\qquad  \cdot {\det}^{1/2}(g)\, d\vec{y}   \left(1 +\frac{t}{4} 
R_{\mu \eta}^{\doublegap \nu
  \pi}\dx^\mu_x\dx^\eta_x \oppartialx{\nu}
\oppartialx{\pi}  + O(t^{3/2})\right) \alpha(0,\dx_x)
\end{align*}
where we have applied Eqs.~\eqref{eq:bint-pt}
and~\eqref{eq:bint-iota}.  
Now, if $f$ is a smooth function on $\RR^{2m}$, then \eqa{flat-1}{
  \int H(\vec{y};t) f(\vec{y}) \, dy^1 \cdots dy^{2m} &= \int (2\pi
  t)^{-m} \exp\left[-\frac{1}{2t} |\vec{y}|^2\right] f(\vec{y}) \, dy^1 \cdots dy^{2m} \nonumber\\
  & =
  f(0) - t\frac{(\Delta f)(0)}{2} + O(t^2) } \noindent where $\Delta
f= -\delta^{\mu \nu} \partial_\mu \partial_\nu f.$ 
Since, according to Eq.~\eqref{eq:rndet},  $\left[1 -
  \frac{1}{6} \Ricci_{\sigma \tau} y^\sigma y^\tau\right]
{\det}^{1/2}(g) = 1 + O(|\vec{y}|^3),$ Eq.~\eqref{eq:flat-1} implies the term linear in $t$ coming from the integral over $\RR^{2m}$ is just $-\frac{1}{2}\Delta \left[f(\vec{y})\left(1+ \frac{1}{2} R_{\pi \eta \sigma}^{\triplegap \tau}y^\pi
  \dx_0^\eta y^\sigma \oppartialx{\tau}\right)\right]_{\vec{y} = \vec{0}}$. That is,
\begin{align*}
\MQO(t) \alpha &= f(0)\left(1+ \frac{t}{2} \Ricci_\eta^\pi 
{\dx}_0^\eta \oppartialx{\pi}  +\frac{t}{4} 
R_{\mu \eta}^{\doublegap \nu
  \pi}{\dx}^\mu_0{\dx}^\eta_0 \oppartialx{\nu}
\oppartialx{\pi}\right)\alpha(0, \dx_x)\\
&\qquad + \frac{t}{2}\left(\delta^{\mu \nu} \partial_\mu \partial_\nu
  f\right)(0) \alpha(0, \dx_x) + O(t^{3/2}).
\end{align*}
Thus the required $t$-derivative is
\begin{align*}
\lim_{t\to 0} \left(\MQO(t) \alpha -\alpha\right)/t&=\frac{1}{2}\left(\delta^{\mu \nu} \partial_\mu \partial_\nu
  f\right)(0) \alpha(0, \dx_x)  \\
& \qquad + \left(\frac{1}{2}\Ricci_\eta^\pi 
{\dx}_0^\eta \oppartialx{\pi}  +\frac{1}{4} 
R_{\mu \eta}^{\doublegap \nu
  \pi}{\dx}^\mu_0{\dx}^\eta_0 \oppartialx{\nu}
\oppartialx{\pi}\right) f(0) \alpha(0, \dx_x).
\end{align*}
On the other hand $\Nabla_\mu \alpha=0$ since it is covariantly
constant, so in Riemann normal coordinates, with the derivatives
acting at $0$, the right-hand side of Assumption~2 is
\begin{eqnarray*}
\lefteqn{-\frac{\Delta}{2} \alpha = -\frac{1}{2} 
\Delta_0 f(x) \alpha(0, \dx_x)= } \\
&&
\frac{1}{2} \left(\delta^{\mu \nu} \partial_\mu \partial_\nu
  f\right)(0) \alpha(0, \dx_x) +\frac{1}{2}  \left(
\Ricci_\eta^\pi {\dx}^\eta \oppartialx{\pi} + \frac{1}{2} R_{\mu \eta}^{\doublegap \nu
  \pi}{\dx}^\mu{\dx}^\eta \oppartialx{\nu}
\oppartialx{\pi}\right) f(0) \alpha(0, \dx_x).\end{eqnarray*}

Assumption~3 is an analogous but simpler calculation where we consider
$\int K(x,y;t) \alpha(x) dx $ for a smooth $\alpha$ and require it to
converge to $\alpha(y)$ as $t$ goes to zero. 
\end{proof}
\begin{corollary}\label{par:cr}
For any sequence of partitions  $t_1, t_2, \ldots,
t_n$ such that  $\max_i (t_i) \to 0$ and $\sum_i t_i \to t$ and for any form
$\alpha$ on $M$
\[\lim {\mathfrak{K}}(t_1) {\mathfrak{K}}(t_2) \cdots {\mathfrak{K}}(t_n) \alpha = e^{-t\Delta/2}
\alpha\]
where $\Delta$ is the Laplace-Beltrami operator on forms.  
Moreover, for some such sequence of partitions 
\[\lim K(t_1) * K(t_2) * \cdots * K(t_n) \to K_\Delta(x,y;t)\]
uniformly, where $K_\Delta$ is the heat kernel of $\Delta $ (the
kernel of $e^{-t\Delta/2}$).   
\end{corollary}
\begin{remark}
Thus the
approximation $\MQ(x,y; t_1, \ldots, t_n)$ to the kernel of the time
evolution operator for supersymmetric quantum mechanics converges to
the heat kernel for $\Delta$ in the large partition limit.
\end{remark}

\def\cprime{$'$}

\end{document}